\documentclass[runningheads]{llncs}
\usepackage{cite}
\usepackage{amsmath,amssymb,amsfonts}
\usepackage{algorithmic}
\usepackage{graphicx}
\usepackage{textcomp}
\usepackage{xcolor}
\usepackage{hyperref}
\usepackage{setspace}
\newcommand\etal{\emph{et al.}}
\begin{document}
\title{MIDGET: Music Conditioned 3D Dance Generation}
%
%
\author{Jinwu Wang\inst{1}\orcidID{0009-0005-9817-7043} \and
Wei Mao\inst{1}\orcidID{0000-0002-8876-8983} \and
Miaomiao Liu\inst{1}\orcidID{0000-0001-6485-3510}}
\authorrunning{F. Author et al.}
%
\institute{Australian National University, Canberra ACT 2601, Australia}
\maketitle              
\begin{abstract}
In this paper, we introduce a \textbf{M}us\textbf{I}c conditioned 3D \textbf{D}ance \textbf{GE}nera\textbf{T}ion model, named \textbf{MIDGET} based on Dance motion Vector Quantised Variational AutoEncoder (VQ-VAE) model and Motion Generative Pre-Training (GPT) model to generate vibrant and high-quality dances that match the music rhythm. To tackle challenges in the field, we introduce three new components: 1) a pre-trained memory codebook based on the Motion VQ-VAE model to store different human pose codes, 2) employing Motion GPT model to generate pose codes with music and motion Encoders, 3) a simple framework for music feature extraction. We compare with existing state-of-the-art models and perform ablation experiments on AIST++, the largest publicly available music-dance dataset. Experiments demonstrate that our proposed framework achieves state-of-the-art performance on motion quality and its alignment with the music. 

\end{abstract}
\section{Introduction}

\begin{figure}[th!]
  \centering
  \includegraphics[width=12cm]{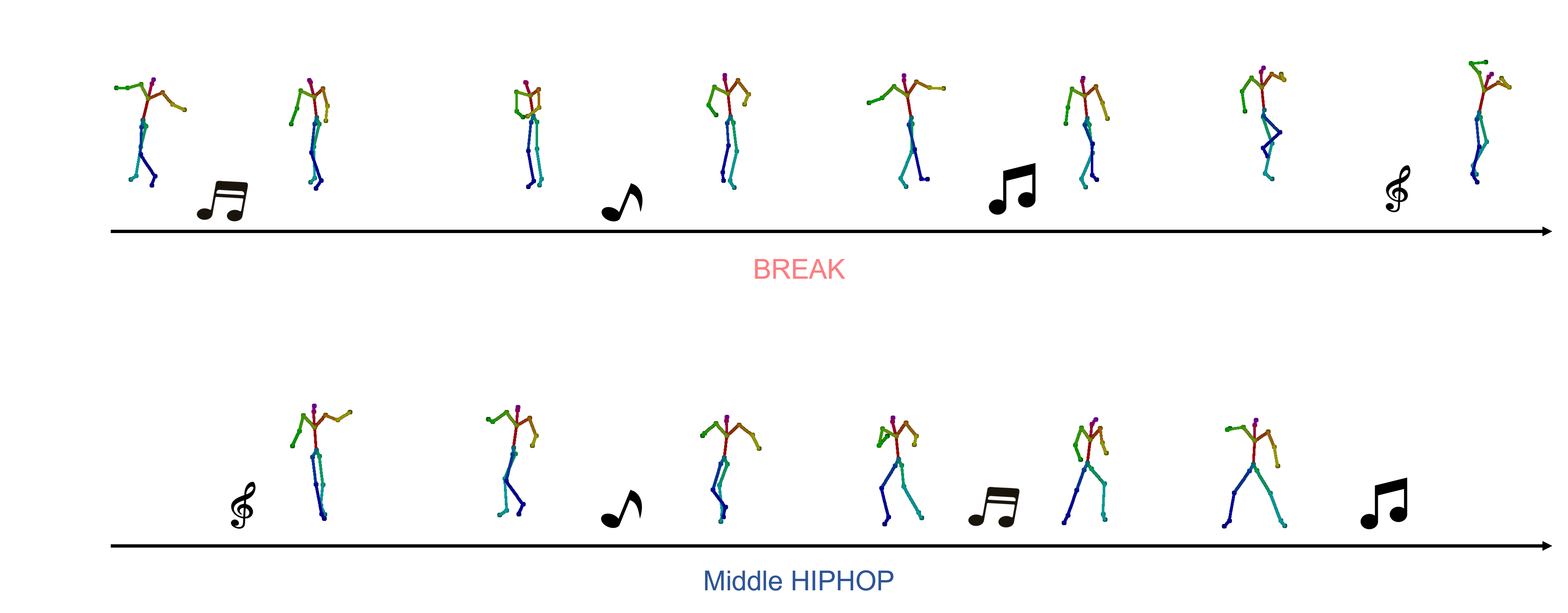}
  \caption{\textbf{Dance examples generated by our proposed method.} Qualitative human motion generation samples based on our MIDGET model can be found at \href{https://youtube.com/playlist?list=PLFUM19_jtCvR7ThXF6dyQCaGX3hmj416Z}{YouTube}. \label{fig:1}}
\end{figure}%
Dance is an art form that expresses emotions and culture through-composed motion patterns. \cite{range}. Moreover, music as the carrier of dance is an indispensable element in dance performance. All dances always have some strong correlations with the music rhythm\cite{range, time_dance}. Music-based human motion generation drives many applications like choreograph and virtual character animation~\cite{bailando}.

One of the main challenges of this task is to ensure the consistency between the music rhythm and the generated motion e.g., the alignment of music beats and motion. 
Previous methods e.g., EDGE~\cite{edge} learn it implicitly from the data which   
cannot ensure the motion-music beat alignment leading to artifacts like freezing motion. Although other methods such as Bailando~\cite{bailando} try to explicitly encourage the music-motion beat alignment, their actor-critic learning strategy is very unstable during train. 
By contrast, in this paper, we propose a gradient copying strategy which enable us to directly train the motion model with beat alignment loss in a gradient descent manner. Furthermore, other than Bailando~\cite{bailando} which directly downsample the music feature to match the human motion, we propose to learn a music feature extractor which is proved to be more effective. In particular, building upon the Bailando~\cite{bailando}, we introduce MusIc conditioned 3D Dance GEneraTion (MIDGET), an end-to-end generative model for generating motion that match the beats of the music.

The main contributions of our work can be summarized as follows:
\begin{itemize}
    
    \item We introduce a gradient copying strategy which enables us to train the motion generator with music alignment score directly. 
    \item We propose a simple yet effective music feature extractor improves recognition and analysis performed on music information with few additional parameters.
\end{itemize}
\section{Related Work}
The field of music to dance generation research generates dance motions for the traditional music of a specific style or genre through machine learning ideas\cite{Min, L2D, graph}. This area has been studied for many years, and various techniques have been used to implement dance motions generation.

\noindent{\bf Traditional Music-to-Dance Models}
Lee \textit{et al.}\cite{Min} used a clustering model to group musically similar segments into a cluster and then used the dance motion sequences corresponding to the same clusters as the generated results. This method determines dance motions by classifying music only and needs more diversity and accuracy of dance motions. Ofli \textit{et al.}\cite{L2D} proposed using four statistical machine learning models to obtain the mapping relationship from music to capture the diversity in dance performances and the dependence on musical segments, and dance figure models are used to model the motion of each dance character to capture the changes in the performance style of a particular dance character.

In addition, 
Shiratori \textit{et al.}\cite{graph} uses a graph-based approach to cut motion clips from existing data into individual nodes and stitch them together to synthesize new motions based on appropriate musical features. However, this clip-based scheme unitizes the motions. So that only a fixed rhythm, tempo, and variety of motions can exist. 

\noindent{\bf Music-Conditioned 2D Human Dance Generation} Most currently available works have studied the 2D choreographic generation of music, Lee ~\etal\cite{lee} realized that dance is multi-modal and that various subsequent pose motions are equally possible at any moment. The Generative model of music generative dance is proposed for the first time to decompose dance into a series of basic dance units. In the synthesis phase, the model learns to choreograph a dance by seamlessly organizing multiple basic dance motions based on the input music. Li ~\etal\cite{li2017auto} connects the predicted output of the network itself with its future input stream by using the training mechanism of an automatic conditional recurrent neural network\cite{rnn}. Qi ~\etal\cite{driven} proposed a sequence-to-sequence learning architecture that leverages LSTM\cite{lstm} and Self-Attention mechanism for dance generation based on music. 

Huang~\etal\cite{huang2020dance} introduces Transformer to realize 2D dance generation under music conditions and uses the local self-attention to efficiently handle long sequences of music features. In addition, Huang \textit{et al.} \cite{huang2020dance} proposes a dynamic automatic conditional training method to mitigate the error accumulation of auto-regressive models in long motion sequence generation.

\noindent{\bf Music-Conditioned 3D Human Dance Generation} The application scenario of 2D dance motion is minimal. It only contains flat information and thus needs 3D spatial three-dimensional details. Research in the Music-Conditioned 3D Human Dance Generation field has emerged. Li \textit{et al.}\cite{fact} and Li \textit{et al.}\cite{danceformer} allows the models to extract and learn audio features by stacking different forms of Transformers. Siyao \textit{et al.}\cite{bailando} used the idea of VQ-VAE\cite{vqvae, razavi2019generating} to encode and quantize the spatial standard dance motions into a limited codebook, which can store different dance motions in a codebook so that a more professional, reasonable, and coherent dance motion can be used when generating a sequence. Each sequence inside the codebook represents a unique dance pose, and these dances have contextual semantic information.

 Recently, diffusion-based approaches\cite{diffusion, edge,magic} have excelled in generating musically conditioned motion to create realistic and physically plausible dance motions based on input music. Tseng \textit{et al.}\cite{edge} proposed Editable Dance Generation (EDGE), a transformer-based diffusion model paired with the powerful music feature extractor Jukebox\cite{jukebox, edge}. It was demonstrated that this unique diffusion-based approach gives powerful editing capabilities well suited to dance, including joint modulation and intermediate processing.

\section{Method}
\begin{figure}[t]
  \centering
  \includegraphics[width=12cm]{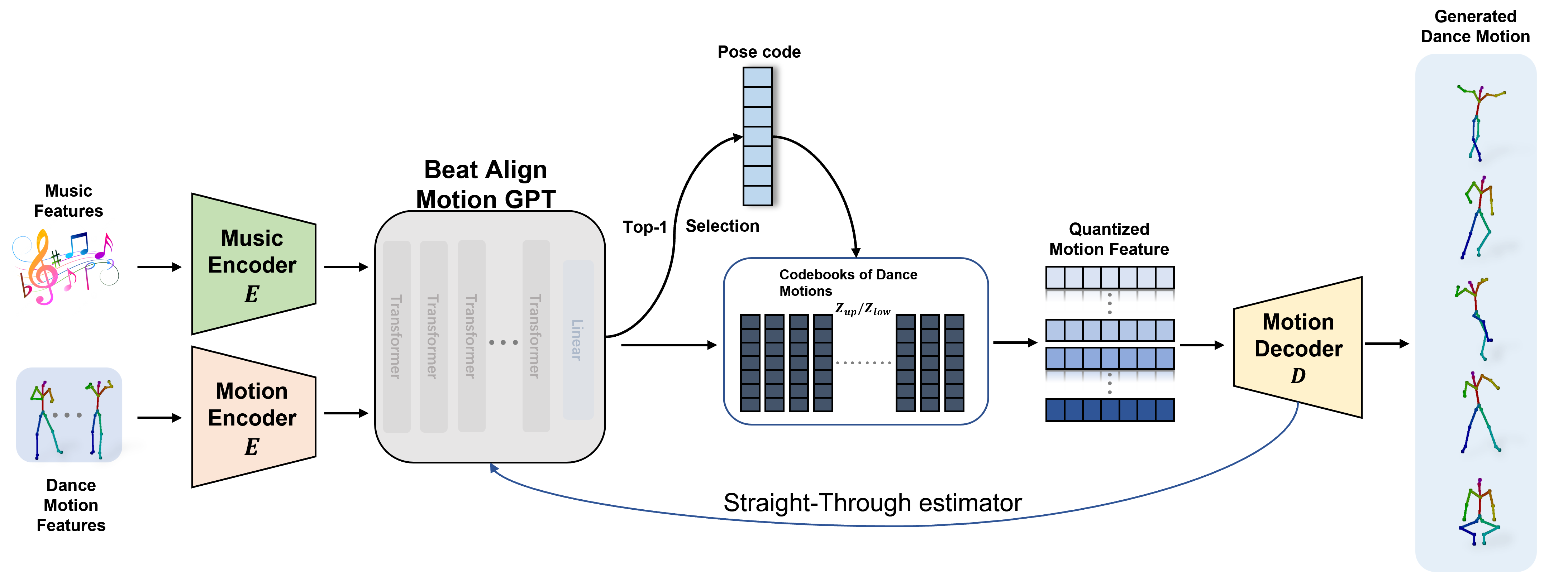}
  \caption{\textbf{Overview of the MIDGET Model.} Given a piece of music and its corresponding dance instructions, MIDGET can generate corresponding high-quality and smooth dance sequences. \label{fig:model}}
\end{figure}%

An overview of the proposed music-to-dance generation framework, "Music Conditioned 3D Dance Generation (MIDGE)", is shown in Fig.~\ref{fig:model}. We first quantize the dance motions using the VQ-VAE~\cite{vqvae} resulting a finite-length codebook $\mathcal{Z}=\{z_i\}_{i=1}^{K}$ where each code $z_i \in \mathbb{R}^{C}$ is a vector of size $C$. 
Each motion sequence is then represented as a sequence of codes from the codebook.  
Instead of directly downsampling the music feature as in~\cite{bailando}, we adopt a 1-D convolutional network to learn such downsampling process which is proved to be more effective.
Conditioning on the music feature and a seed motion, we leverage the motion GPT~\cite{bailando} to obtain the code sequences for both upper body and lower body. Such code sequences can be easily decoded as human motion via the pretrained VQ-VAE. 
We provide the details for each module below.

\subsection{3D Dance Motion VQ-VAE}\label{vq_vae}

In order to effectively capture and extract the unique style of dance motion codes and enable them to be reconstructed into corresponding dance motion sequences with actual physical meaning. 
Therefore, we use 1-D temporal convolutional network to encode dance motion sequences $P\in \mathbb{R}^{T\times(J\times 3)}$ into pose codes $Q\in \mathbb{R}^{T'\times C}$ 
, where $T$ denotes the time duration of the original dance motion sequence and $J$ means the number of joint, and $C$ is the feature dimension, and $T'=T/d$ is the motion code length of the pose codes, and $d$ is the temporal sampling rate. 

\begin{figure}[th!]
    \centering
    \includegraphics[width=12cm]{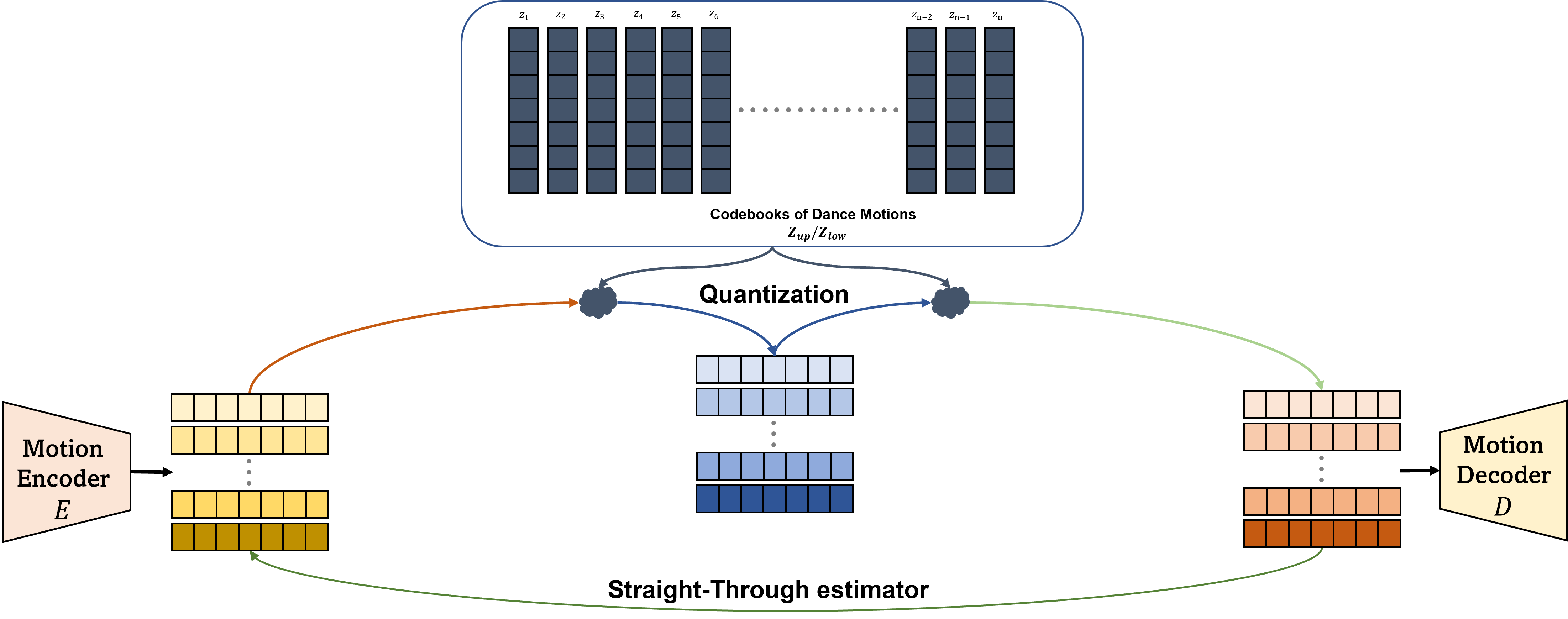}
    \caption{\textbf{3D Dance Motion VQ-VAE.} The main purpose of the VQ-VAE model is to obtain codebooks containing diverse quantified dance motion sequences. Learnable encoder and decoder to quantify features and reconstruct target poses. \label{fig:vqvae}}
\end{figure}%

\textbf{Codebook of Dance Poses.} 
The VQ-VAE codebook\cite{vqvae} is a vector collection that maps continuous high-dimensional data into discrete ones. Each of these vectors is called a code. 
By transforming continuous data into discrete representations, the codebook is able to store and process data in lower dimensions, enabling efficient data compression. In this paper, we mapped the output of the motion encoder module to the  nearest neighbour vector in the codebook to discretize the continuous high-dimensional data into a low-dimensional code. In this scheme, we analyze the output $e_i\in\mathbb{R}^{C}$ of the encoder
and map it to the nearest discrete pose code in the codebook:
\begin{equation}\label{nearest}
    \begin{aligned}
            q_i =\text{arg}\min_{z_j\in \mathcal{Z}}||e_i-z_j||_2\;,
    \end{aligned}
\end{equation}
where $q_i\in\mathbb{R}^{C}$ is the $i$-th row of $Q$ and $i \in \{1,2,\cdots,T'\}$.
Finally, The motion decoder is defined as a 1-D temporal deconvolutional network to reconstruct the corresponding dance pose sequences $\hat{P}\in \mathbb{R}^{T\times{J\times3}}$ from the latent codes.

\textbf{The training process of Motion VQ-VAE.} 
With this gradient conduction, we can achieve the training of the Motion Encoder $E$, Decoder $D$ and codebook $\mathcal{Z}$ simultaneously with the following loss function:
\begin{equation}
    \begin{aligned}
        \mathcal{L} = \mathcal{L}_{REC} + \mathcal{L}_{VQ} + \beta\cdot\mathcal{L}_{COM}
    \end{aligned}
\end{equation}
The loss function consists of three terms, where $\mathcal{L}_{REC}$ is the ~\emph{reconstruction~loss} used to assess the quality of dance motions. It measures the difference between the generated and target pose sequences. In particular, we propose to define the reconstruction loss function by focusing on the original 3D coordinate points of the body joints and the velocity and acceleration of the motion sequences. The detailed definition is presented below:
\begin{equation}
    \begin{aligned}
       \mathcal{L}_{REC} = \left\Vert P-\hat{P}\right\Vert_2 + \alpha_1\left\Vert V - \hat{V}\right\Vert_2 + \alpha_2\left\Vert A-\hat{A}\right\Vert_2
    \end{aligned}
\end{equation}
where $P,\hat{P}$ are the ground truth and generated joints of the original 3D points, respectively. $V$ and $A$ represent the motion velocity and acceleration partial derivatives of the 3D joint sequence on time to learn the time-dependent and spatial relationships in the dance motion sequences, thus learning the transition patterns between different motions and the structure of the motion sequences.

The second and third part of $\mathcal{L}$ is the quantification loss with $\mathcal{L}_{VQ}$ and $\mathcal{L}_{COM}$ to constrain the difference between the discretized code and the continuous code. $\mathcal{L}_{VQ}$ using exponential moving averages function to learn latent embedding vectors in the codebook. We use Exponential Moving Averages (EMA)\cite{EMA, vqvae} to update the vectors in the codebook of the VQ-VAE model. 

\begin{equation}\label{code_loss}
    \begin{aligned}
       \mathcal{L}_{VQ} &= \left( e_i, q_i\right)_{EMA}
    \end{aligned}
\end{equation}

$\mathcal{L}_{COM}$ defined in Eq.~\eqref{com_loss} constrains the distance between the encoder output $e_i$ and the decoder input $q_i$ which is reduced to minimise the Euclidean distance between the encoder output and its nearest neighbour quantization centre. The stop gradient\cite{exploring, straightthrough} the nearest neighbour vector $q_{i}$ as a constraint on the vector quantization operation. 


\begin{equation}\label{com_loss}
    \begin{aligned}
       \mathcal{L}_{COM} &= \left\Vert e_i - \text{sg}[q_i]\right\Vert_2
    \end{aligned}
\end{equation}

where the $sg[\cdot]$ denotes "stop gradient". 

\subsection{Music Feature Extraction}\label{music_extractor}
\begin{figure}[th!]
  \centering
  \includegraphics[width=12cm]{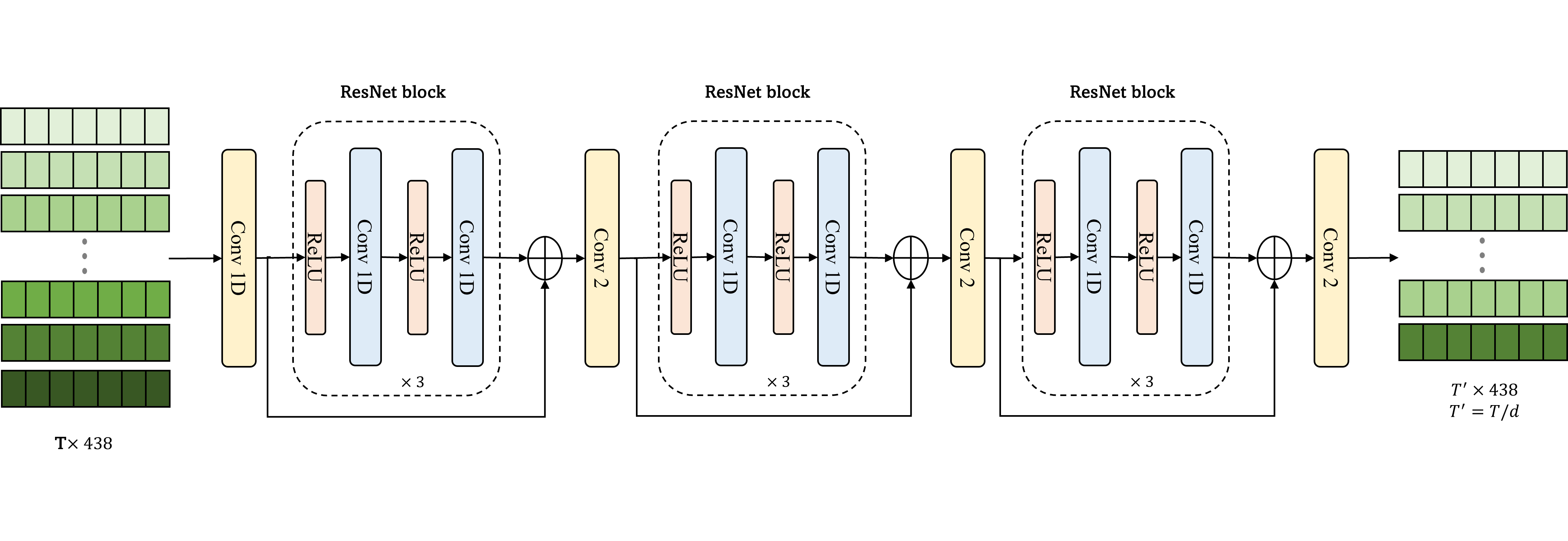}
  \caption{\textbf{The structure of Music Feature Extractor.} Music features are further extracted through one-dimensional convolutional layers and residual connections.\label{fig:extractor}} 
\end{figure}

The Music Feature Extractor is presented in Fig.~\ref{fig:extractor}. The learned music features enable the model to ensure less feature information loss with little additional parameters. The purpose of such feature extractor is to downsample the musical features so as to match the quantized motion codes.

As shown in Fig.~\ref{fig:extractor}, our detailed implementation uses a 1-D convolution and residual structure to downsample the music feature.

\subsection{Motion GPT}\label{motion_gpt}

\begin{figure}[t]
  \centering
  \includegraphics[width=12cm]{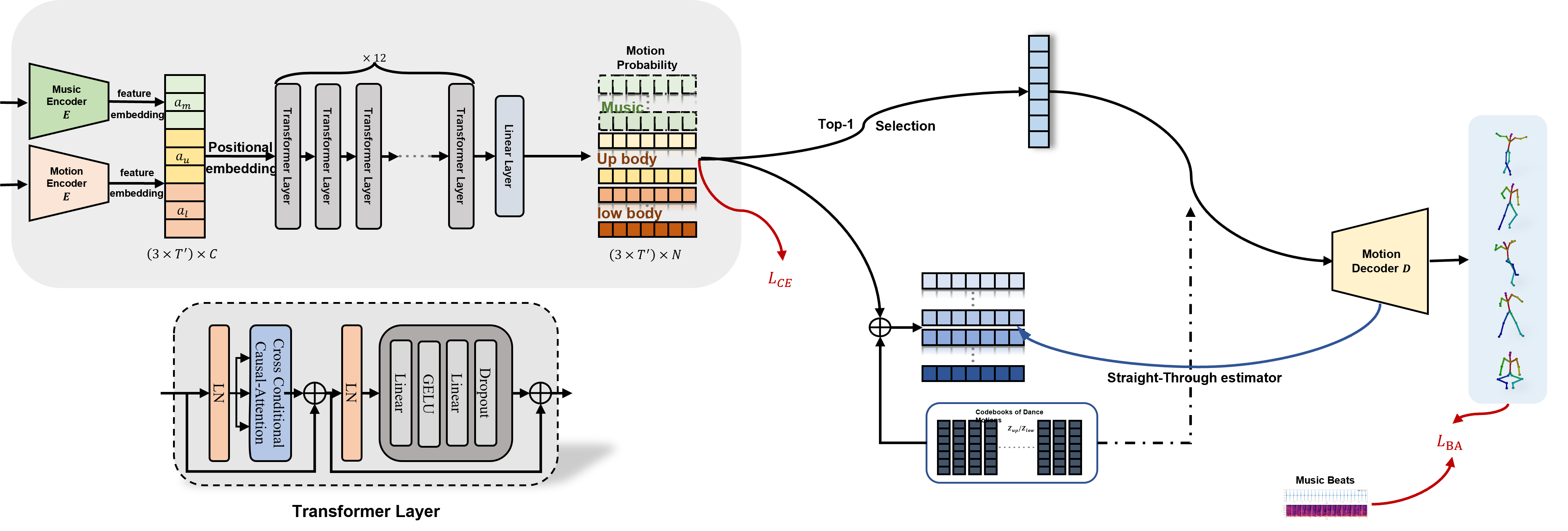}
  \caption{\textbf{Motion GPT Structure.} The GPT model is designed to apply the encoded upper and lower body pose codes $a_u,a_l$ and music features $a_m$ to generate the target future motion probability $p^u, p^l$.\label{fig:gpt_model}}
\end{figure}%

After working with the above 3D Dance Motion VQ-VAE, we can generate unique dance pose clips with physical meaning according to the trained codebook $Z$. Now that we can explore Beats-Aligned Motion GPT, which is a Transformer \cite{transformer, gpt_gnn} based network. The GPT\cite{gpt, gpt_gnn} model must focus on generating future dance motion codes with corresponding music styles under the given music conditions and current starting quantized pose position codes. The Beats-Aligned Motion GPT achieved as Feature Embedding, Positional Embedding and $12$ Transformer Decoders, as shown in Fig.~\ref{fig:gpt_model}.

In order to align the spatial dimensions of the input musical features with the quantized pose position codes. The quantized upper and lower body and music features are subjected to feature embedding operations as $u,l\in \mathbb{R}^{T'\times C_{pose}}$ and $m\in\mathbb{R}^{T'\times C_{music}}$, respectively. Adding a learned position embedding assigns a unique vector representation for each position, which contains the encoding of positional information. Then, the feature vector is propagated through 12 successive Transformers\cite{transformer}, and mapped to motion probability $p\in R^{((3\times T')\times C)}$. When generating future motion sequences, the upper and lower body pose codes  $p^u=p_{T':2T'-1}$, $p^l=p_{2T':3T'-1}$ extracted from motion probability features.


\textbf{Cross-conditional causal attention.}
We adopt "Cross-conditional causal attention" \cite{bailando} which encodes the features across different time series and only allows the information passing from the past. 
It is defined as follows:

\[\mathrm{Attention}(Q,K,V)=\mathrm{Softmax}\left(\frac{QK^T+M}{\sqrt{d}}\right)V\]

where \textbf{d} is the number of channels in the attention layer. \textbf{Q, K, V} denote the query, key, and value from the input feature, and \textbf{$M$} is the mask matrix used to determine the type of attention sub-layers. 

\textbf{3D Dance Motion Generation.}
In order to make the model overall differentiable, we use Straight-Through estimator\cite{straightthrough} to combine VQ-VAE Decoder and codebooks to generate gradient-derivable sequences of actual future motions $\hat{P}_{1:T}$. As shown in Fig.~\ref{fig:gpt_model}, we simulate the generated result of motion GPT as Quantified pose code $\hat{e}_e$, similar to the output feature of Motion Encoder. The method is to perform matrix multiplication of the action encoding probability $p_{1:T'}$ generated by the motion GPT and the codebook $Z$.Because matrix multiplication can be considered as a function, when computing matrix multiplication, the gradient can be computed using the automatic differentiation method, which calculates the partial derivative of the output matrix with respect to the input matrix (see Equations below).


\[\hat{e}_e = p_{1:T'}\cdot Z,\;\; \hat{e}_d = \text{arg}\min_{Z} ||p_{1:T'}||, \;\; \hat{e}_d  = \hat{e}_e + sg[\hat{e}_d - \hat{e}_e]\]

On the contrary, the forward propagation process in the motion GPT generation stage, we perform top-1 selection (select the index with the highest probability) on the coding probability $p_{1:T'}$ of Motion GPT output and match its index to the corresponding position in Codebook and obtain Quantized features $\hat{e_k}$. Finally, the pre-trained VQ-VAE Decoder decodes $\hat{e_k}$ to obtain the future dance motion sequence.

\textbf{The training of Motion GPT.} 
First, since the output of Motion GPT itself is a set of N action code probabilities, which cannot reflect the natural dance motions anyway, we use the upper and lower body motion features from the output of VQ-VAE Encoder as the input of Motion GPT and also as the target $a$ for computing the Mean Squared Error(MSE) Loss on motion code probability $p$:

\begin{equation}
    \begin{aligned}
        \mathcal{L}_{CE} &= \frac{1}{T'}\sum_{t=1}^{T'}\sum_{b=u,l}\left\Vert p_{t}^{b} - a_{t}^{b}\right\Vert_2
    \end{aligned}
\end{equation}
where $p_{t}^{b}$ is the motion code probability, and $a_{t}^{b}$ is the target sequence of VQ-VAE Encoder outputs. Due to the generation mode of the upper and lower half body pose separation, the results of two different loss functions must be combined.

We design Beat Align Loss as Eq.~\eqref{ba_loss} to allow the model to produce more accurate dance motions with music rhythm.

\begin{equation}\label{ba_loss}
    \begin{aligned}
        \mathcal{L}_{BA_{l2}} &= \left\|B_{d} - B_{m}\right\|_2\\
    \end{aligned}
\end{equation}

where $B_d$ is the dance motion beats and $B_m$ is the music beats. We identify dance beats by the difference between the physical position of the action in the front and back frames and use a Gaussian filter to calculate the probability of dance beats. The smaller the distance between the front and back frames, the higher the probability that the frame is a dance beat, as follows:

\begin{equation}\label{B_d}
    \begin{aligned}
        B_{d} &= \exp{\left( -\frac{\|\hat{x}_{(0:T-1)} - \hat{x}_{(1:T)} \|_2} {\sigma^2}\right)}\\
    \end{aligned}
\end{equation}

where $\hat{x}\in \mathbb{R}^{(T\times 24\times 3)}$ is the prediction of the motion sequence. The probability of motion beats is evaluated by calculating the difference between the motion speed of the two nearly frames.

\section{Experimental Result}

Our proposed method is evaluated on the AIST++ dataset published by \cite{fact}, containing 1,408 3D human dance motion sequences represented as joint rotations and root trajectories. 
All dance motions are paired with their corresponding 60 music clips. 

\noindent{\bf Implementation Details.} We train the MIDGET model as a two-stage task with the Adam \cite{adam} optimiser. The model framework is trained on an NVIDIA RTX 4090 GPU for 24 hours with a batch size of $64$.


In our experiments, we separate dance and music into a 240-frame format. The size of the trainable upper and lower body dance memory codebooks in the VQ-VAE model is N=512 dimensions. The downsampling rate in the VQ-VAE Encoder and Music Encoder sections are $8$, so the dimensionality of the quantized codebooks also results from downsampling $Z_{up}, Z_{low}\in \mathbb{R}^{30\times 512}$. 

\noindent{\bf Evaluation Metrics and Baslines} We adopt the $ Fr\acute{e}chet~Inception~Distances$ (FID) \cite{fid, fact, bailando} to evaluate our method by following existing works. 
We similarly measure diversity by calculating the average Euclidean distance between different dance motions in the geometric and kinematic feature space. 
Finally, using the Beat Align score \cite{fact, bailando, edge} and Beat Consistency score \cite{danceformer} to measure the relevance between music and dance motions. Precisely, the explanation and equations are as follows:
\begin{equation}\label{ba}
    \small
    \begin{aligned}
        BA &= \frac{1}{|B^m|}\sum_{t^m\in B^m}\exp{\left\{\frac{\min_{\forall t^d\in B^d}||t^d-t^m||_2}{2\sigma^2}\right\}}\\
        BC &= \frac{1}{|B^d|}\sum_{t^d\in B^d}\exp{\left\{\frac{\min_{\forall t^m\in B^m}||t^d-t^m||_2}{2\sigma^2}\right\}}
    \end{aligned}
\end{equation}

where ${t_i^m}$ is the music beat in $B^m$, ${t_i^d}$ is the dance motion beat in $B^d$, and $\sigma$ is the normalized parameter set to 3 simultaneously. 
\subsection{Results.} 
We compare the performance of our proposed model with the state-of-the-art methods including FACT \cite{fact}, Bailando \cite{bailando} and EDGE \cite{edge}. Table \ref{comparison} shows that our proposed model consistently outperforms Bailando\cite{bailando} in all evaluations under the same underlying model framework and avoids action freezing problem. 
Specifically, our approach improves by 6.8\% and 22.3\% in $FID_k$ and $FID_g$, respectively, for evaluating the physical features of dance. 

Moreover, \textbf{MIDGET} can generate more diverse dances, as reflected in the improvement of $Div_k$ and $Div_g$ metrics by 10.3\% and 1.6\%. Finally, the Beats Align Score achieves 10.5\% higher, suggesting that generated dance motions are better aligned with music beats. In addition, compared with Li~\etal\cite{li2017auto} and FACT\cite{fact}, our proposed model has a significant advantage in all evaluations.
Although \textbf{MIDGET} does not outperform EDGE\cite{edge}, it has lighter structure.
\begin{table*}[htbp]
    \caption{\textbf{Evaluation of Existing models on AIST++ dataset.} Compared to the Ground Truth and three recent state-of-the-art methods, }\label{comparison}
    \centering
    \setlength{\tabcolsep}{3.5pt}
    \small
    \begin{tabular}{|c|cccccc|}
        \hline
        Method & $\text{FID}_k\downarrow $ & $\text{FID}_g\downarrow $ & $\text{DIV}_k\uparrow $ & $\text{DIV}_g\uparrow $ & $\text{BA Score}\uparrow $ & $\text{BC Score}\uparrow $\\
        \hline 
        Ground Truth & 17.10  & 10.60 &  8.19 & 7.45 & 0.2374 & 0.2083\\
        \hline
        Li \textit{et al.}\cite{li2017auto} & 86.43  & 43.46 &  6.85 & 7.45 & 0.201 & 0.203\\
        FACT\cite{fact} & 35.35  & 15.55 &  5.94 & 6.18 & 0.221 & 0.203\\
        Bailando\cite{bailando} & 30.43  & 11.42 &  7.83 & 6.34 & 0.233 & 0.208\\
        EDGE\cite{edge} & -  & - &  10.03 & 6.67 & 0.263 & 0.210\\
        \textbf{MIDGET}(Ours) & 28.51  & 8.87 & 8.64 &  6.44 & 0.254 & 0.212\\
        \hline 
    \end{tabular}
\end{table*}


However, although our model does not outperform the EDGE model \cite{edge}, \cite{edge}, 
requires to pre-train a model, namely Jukebox \cite{jukebox}  to extract features for all music sequences for up to three days, which is extremely time consuming. We achieved similar performance using fewer resources than the EDGE model with a light-weight feature extractor.

\subsection{Ablation Studies}\label{ab_st}
We perform the ablation study on 3D Motion VQ-VAE module, the music feature extractor and the Motion GPT estimator. The results are shown in Table~\ref{ablation}.
\begin{table}[htbp]
    \centering
    \small
    \caption{\textbf{Ablation Study on AIST++ testset.} The experiments cover Motion VQ-VAE, Music Feature Extractor, Best Match, and generation strategies.}\label{ablation}
    \setlength{\tabcolsep}{10pt}
    \begin{tabular}{|c|cccc|}
        \hline 
        Method & $\text{FID}_k\downarrow $ & $\text{FID}_g\downarrow $  & $\text{BA Score}\uparrow $ & $\text{BC Score}\uparrow $\\
        \hline 
        w/o. VQ-VAE & 120.45  & 41.24  & 0.259 & 0.200\\
        w/o. upper/lower & 64.07  & 14.82 & 0.246 & 0.203\\
        w/o Extractor & 30.43  & 11.42 & 0.243 & 0.203\\
        w/o $\mathcal{L}_{BA}$ & 29.10  & 7.25  & 0.239 & 0.205\\
        Single Generated  & 29.34  & 9.84 & 0.288 & 0.209\\
        \hline 
        \textbf{MIDGET} & 28.51  & 8.87 & 0.254 & 0.212\\
        \hline 
    \end{tabular}
\end{table}

\textbf{Motion VQ-VAE.} When we remove the quantized codebook in VQ-VAE\cite{vqvae, bailando} to store the dance motion features, the GPT model cannot generate realistic and physically meaningful body poses while the $\text{FID}_k$ and $\text{FID}_g$ becomes very high. Moreover, we also analyzed whether it is necessary to separate the upper and lower body for independent codebook training. Thus, we just adopt the entire body action sequence as input to train VQ-VAE and Motion GPT. The obtained $FID_k$ and $FID_g$ are 124.7\% and 66.9\% which is worse than estimating the code for the upper and lower body, respectively.


\textbf{Music Feature Extractor.} 
We adopt Beat Align Score to evaluate the effectiveness of our proposed Music Feature Extractor. Results in Table \ref{ablation} show that
the BA Score obtained with Music Feature Extractor is higher than the downsampling of music features alone (4.53\% improvement). This indicates that our proposed Music Feature Extractor can extract more effective music features. 

\begin{figure}[ht!]
  \centering
  \includegraphics[width=9cm]{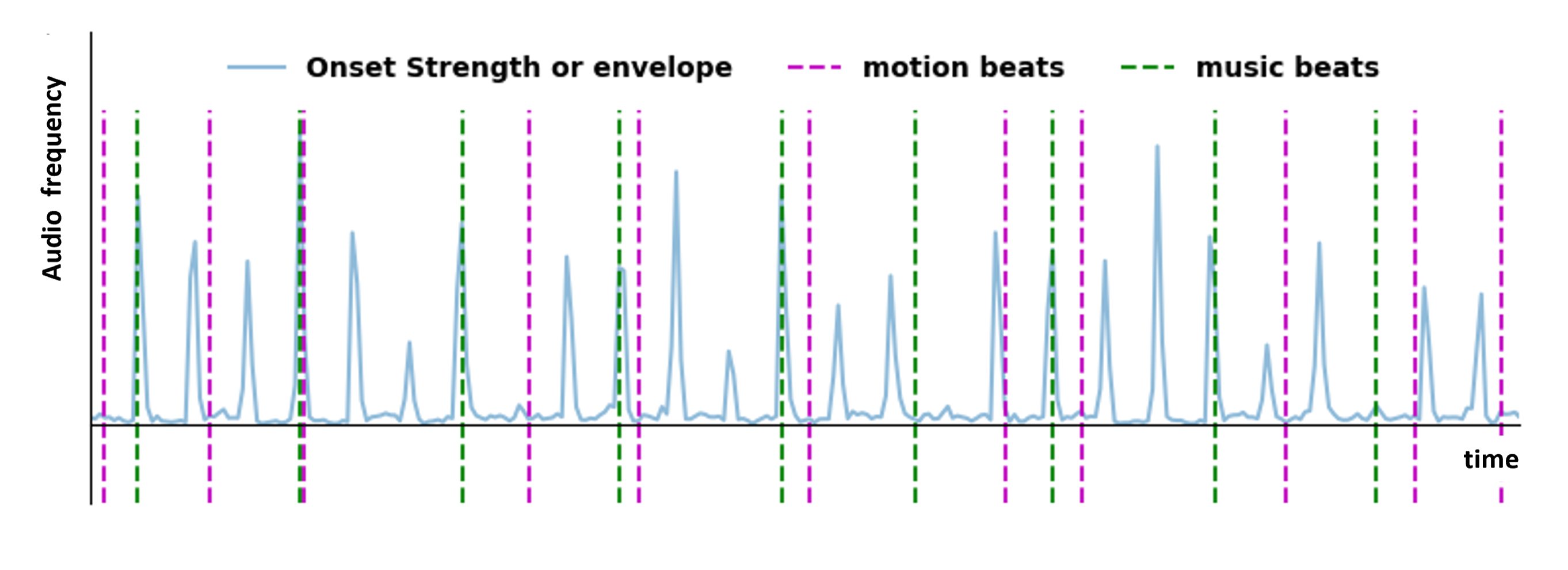}
  \caption{\textbf{Beats alignment between Music and Generated Dance.} Visualize the alignment of dance beats (purple dashed lines) and music beats (green dashed lines).}\label{fig:com_ba}
\end{figure}

\textbf{Beats Match.} 
With our introduced Beat Align Loss, the Motion GPT significantly improves the alignment between actions and music beats, which has increased by 20\% and 10\% on BA Score and consistency scores (BC Score), respectively. 
The detailed beat alignment effect is shown in Figure \ref{fig:com_ba}. 

\textbf{Generation Strategy.} 
We evaluate the generation strategy by just predicting 1-dimensional pose code ($8$ frames) only using Motion GPT during training. The long sequence is generated in an auto-regressive manner. While the generated sequence has a higher BA Score (see Table \ref{ablation}), the results suffer from severe motion freeze issue.

\section{Conclusion}
We proposed MIDGET,  which can generate realistic, and smooth long-sequence dance motions. We have introduced a Beat Align Loss and the Straight-Through Estimator to achieve the end-to-end training, and a simple music feature extractor to improve music-feature learning. The method can largely solve the motion freeze issue for long-sequence generation evidenced by the experiments on AIST++ dataset and achieves superior performance.

%
%

%
%
%
%

\bibliographystyle{splncs04}
\bibliography{bib}





\end{document}